%\mag=\magstephalf
\mag=\magstep1
\pageno=1
\input amstex
%\baselineskip = 0.8 true cm
\documentstyle{amsppt}
\TagsOnRight
\interlinepenalty=1000
%\hsize=6.5truein
%\voffset=0.3truein
%\hoffset=0.2truein
%\vsize =9.8truein
\NoRunningHeads
%\pagewidth{14cm}
%\pageheight{21.5cm}
%\vcorrection{-1.0cm}
%\hcorrection{-1.2cm}

\pagewidth{16. truecm}
\pageheight{23.0 truecm}
%\vcorrection{-1.0cm}
%\hcorrection{-1.2cm}
%\advance\vsize by -\voffset
%\advance\hsize by -\voffset
%\baselineskip = 1.0 true cm
\nologo

\font\twobf=cmbx12

\define \Bone{1}
\define \Btwo{2}
\define \Bthree{3}
\define \BEL{4}
\define \Can{5}
\define \Gran{6}
\define \Kan{7}
\define \Ma{8}
\define \Oone{9}
\define \Otwo{10}
\define \Othree{11}
\define \ORGT{12}
\define \Take{13}
\define \Web{14}
%\define \WW{15}

%\define \tvskip{\vskip 0.5 cm}
\define \tvskip{\vskip 1.0 cm}
\define\ce#1{\lceil#1\rceil}
\define\dg#1{(d^{\circ}\geq#1)}
\define\Dg#1#2{(d^{\circ}(#1)\geq#2)}
\define\dint{\dsize\int}
\def\fp{\flushpar}

\define\s#1{\sigma_{#1}}
\define\tp#1{\negthinspace\left.\ ^t#1\right.}
\define\mrm#1{\text{\rm#1}}
\define\lr#1{^{\sssize\left(#1\right)}}

\redefine\qed{\hbox{\vrule height6pt width3pt depth0pt}}

{\centerline{\bf{Recursion Relation of Hyperelliptic
Psi-Functions of Genus Two}}}

\author
Shigeki MATSUTANI${}^0$%\footnote
\endauthor
\affil
8-21-1 Higashi-Linkan Sagamihara 228-0811 Japan
\endaffil
 \endtopmatter

\footnotetext{e-mail:RXB01142\@nifty.ne.jp}
\baselineskip = 0.8 true cm

\vskip 0.5 cm
\centerline{\twobf Abstract }\vskip 0.5 cm

A recursion relation of hyperelliptic $\psi$ functions of genus two,
which was derived by  D.G. Cantor
 (J. reine angew. Math. {\bf 447} (1994) 91-145), is studied.
As Cantor's approach is algebraic,
another derivation is presented
as a natural extension of the analytic derivation of the recursion
relation of the elliptic $\psi$ function.

\vskip 0.5 cm

{\centerline{\bf{
}}}

\newpage
\baselineskip = 0.8 true cm

\vskip 0.5 cm
\centerline{\twobf \S 1. Introduction }\vskip 0.5 cm

For the case of elliptic curves, the elliptic
$\psi$-function is
defined by \cite{\Take, \Web},
  $$
 \psi_n(u)=
   \frac{\sigma(nu)}{\sigma(u)^{n^2}}.
  \tag 1-1
  $$
Using the addition formula of elliptic $\wp$ functions,
$$
        -(\wp(z) - \wp(u)) =
\frac{\sigma(z+u)\sigma(z-u)}{[\sigma(z)\sigma(u)]^2},
         \tag 1-2
$$
we have a recursion relation \cite{\Take, \Web},
$$
 \psi_{n+m} \psi_{m-n}
=\left| \matrix
 \psi_{m-1}\psi_n & \psi_{m}\psi_{n+1} \\
\psi_{m}\psi_{n-1} & \psi_{m+1}\psi_{n}
\endmatrix \right|. \tag  1-3
$$
Eq.(1-3) is proved as follows.
The addition formula (1-2) becomes,
$$
        -(\wp(nu) - \wp(u) ) =
\frac{\psi_{n+1}(u) \psi_{n-1}(u) }
           {{ \psi_{n}(u)^2  }},\tag 1-4
$$
and
$$
        -(\wp(mu) - \wp(nu))
= \frac{\psi_{n+m}(u) \psi_{m-n}(u) }
        {{ \psi_{m}(u)^2 \psi_{n}(u)^2  }}.
\tag 1-5
$$
Noting
$$
[-\wp(mu) + \wp(nu)] + [\wp(mu) - \wp(u)]-[\wp(nu) - \wp(u)]=0,
\tag 1-6
$$ we
have the relation (1-3).
The formula (1-3) is a recursion relation of the elliptic
$\psi$-function.

In this article
we will consider an extension of these properties Eqs.(1-1)-(1-6)
to those of a hyperelliptic curve $C$ of genus two
given by an affine equation
$y^2=x^5 + \lambda_4 x^4 + \lambda_3 x^3 + \lambda_2 x^2 $
$ + \lambda_1 x + \lambda_0$ for certain complex numbers $\lambda$'s.

Recently Cantor found a generalization of the relation (1-3) of
 hyperelliptic $\psi$ function of genus two \cite{\Can},
$$
 \psi_2^2 \psi_m\psi_n \psi_{n+m} \psi_{m-n}
=\left| \matrix
 \psi_{m-2}\psi_n & \psi_{m-1}\psi_{n+1} & \psi_m\psi_{n+2}\\
\psi_{m-1}\psi_{n-1} & \psi_m\psi_n & \psi_{m+1}\psi_{n+1}\\
\psi_m\psi_{n-2} &\psi_{m+1}\psi_{n-1}& \psi_{m+2}\psi_n
\endmatrix \right| , \tag 1-7
$$
by using Hankel determinant and Pad\'e approximations.
Here the hyperelliptic $\psi$ function of genus two is defined as
a polynomial $\psi_m$ of $x$ and $y$,  whose zeros $P$
in the curve $C$ are necessary and sufficient condition for
the element $m \cdot P$ in the related Jacobian
(Jacobi variety) $\Cal J$ to stay
in the  curve $C$ embedded in $\Cal J$ again.
Due to the definition, $\psi$ is an object of concern in
number theory \cite{\Can, \Oone-\Othree}.
(By primitive consideration, for any hyperelliptic curve
we find that twice of the finite ramified points, {\it i.e.},
zeros of $y$, in the Jacobian
exist in the curve again. In other words,
 $\psi_2$ must be $y$ up to constant factor  \cite{\Oone, \Otwo}.
However as the fact that  $\psi_2$ is proportional to $y$
is obtained by a trivial consideration, Cantor removed it after
scaling $\psi_m$ by $\psi_2$ for the sake of convenience
from the viewpoint of study of cryptography.
In fact in original formula of Eq.(1-7) in \cite{\Can},
the factor $\psi_2^2$ in the left hand side in Eq.(1-7) does not exist.
In this article, we will faithfully follow the definition and
employ convention of \^Onishi \cite{\Oone, \Otwo}
which requires some corrections to ones of \cite{\Can}.)
We employ an analytic expression of $\psi$ function \cite{\Oone, \Otwo},
$$
        \psi_n(u)=\frac{\sigma( n u )}{\sigma_2( u)^{n^2}}, \tag 1-8
$$
where $\sigma$ is the hyperelliptic $\sigma$ function defined in
\S 2, and $u$ is a coordinate of the embedded curve $C$ in
the related Jacobian $J$.

In \cite{\Ma}, I showed that
the relations (1-3) and (1-7) can be regarded as
discrete nonlinear difference equations themselves
and are closely related to the discrete Painlev\'e equation I
\cite{\ORGT},
$$
        \beta_{n+1}\beta_{n-1} = \frac{z}{\beta_{n}}+
 \frac{a}{\beta_{n}^2}, \tag 1-9
$$
and its higher rank analog.
By setting $\beta_n = \psi_{n+1}
\psi_{n-1}/\psi_{n}^2$, Eq.(1-3) is reduced to Eq.(1-9) \cite{\Ma}.
Similarly we obtain several similar equations to Eq.(1-9)
using the recursion relation (1-7).
Thus Eq.(1-7) is a very interesting formula even from the
viewpoint of the study of the nonlinear
 difference equation.
Our motivation of this study is to find the
extension of Eq.(1-9) from the viewpoint.
In order to do, we believe that the investigation of
the mathematical structure of Eq.(1-7) is very important.

Though Eq.(1-7) was obtained by Cantor,
his approach is not out of the line of the derivation of
Eqs.(1-1)-(1-6).
His approach to Eq.(1-7) \cite{\Can} is algebraic
(and  applicable to case of
hyperelliptic curves of any genus).

In this article,
we will give an {\it analytic} derivation of Eq.(1-7) as a simple
extension of elliptic case shown in Eqs.(1-1)-(1-6) in the
meaning of {\it study of special functions}.
(In this article, {\it we have used only the term
"analysis" as a classical meanings or study of special functions
or algebraic functions.})
For the case of genus one,
the left hand side of Eq.(1-3),
$\psi_{m+n}\psi_{m-n}$ comes from
the factor $\sigma(u+v)\sigma(u-v)$ in the right hand
side of Eq.(1-2) due to the definition of  $\psi$ in Eq.(1-1).
The linear terms with respect to $\wp$ in
the addition formula (1-2)
play essential roles and  the relation (1-6)
is reduced to the recursion relation (1-3).

Our plan to prove the formula (1-7) is an extension of
 the idea given in Eq.(1-6).
For the case of the genus two, using the hyperelliptic
$\sigma$ function, hyperelliptic $\psi$ function
is given by Eq.(1-8).
On the other hand, the addition formula is given as
a relation in the Jacobian \cite{\Btwo, \BEL},
$$
\frac{\sigma(u+v)\sigma(u-v)}{\sigma(u)^2\sigma(v)^2}
=-(\wp_{11}(u)-\wp_{11}(v) + \wp_{12}(u)\wp_{22}(v) -
                          \wp_{12}(v)\wp_{22}(u)),
                          \tag 1-10
$$
for the hyperelliptic $\sigma$ and $\wp$ functions.
As in the case of genus one, we will use the role of
the linear terms of $\wp$ in Eq.(1-10) like in Eq.(1-6).

However there are two different points from the case of
genus one. First is that the addition formula (1-10) contains
the second order terms $ \wp_{12}(u)\wp_{22}(v) -
\wp_{12}(v)\wp_{22}(u))$ of $\wp$. Second is that
the hyperelliptic $\psi$ function is defined over
the embedded curve $C$ in the Jacobian $J$ whereas
the the hyperelliptic $\sigma$ and $\wp$ functions
and their relations are defined over the Jacobian $J$.

In order to overcome the difference, we introduce a function,
$$
   \Psi_n :=\frac{\sigma( n u )}{\sigma_2( u)^{n^2}},
    \tag 1-11
$$
whose domain is $J$ and consider  lifted quantities of
both sides of the recursion relation (1-7).
We evaluate the parts in terms of the relations of
the $\wp$-functions in the Jacobian $J$. After then, we constrain
the quantities to the embedded curve $C$ in the
Jacobian $J$.
Thus \S 2 devotes the review of the hyperelliptic
functions over the Jacobian $J$ and related
analytic relations as a sense of study of special functions
 following the studies of hyperelliptic functions
\cite{\Bone, \Btwo, \Bthree, \BEL, \Gran, \Oone, \Otwo}.
Using the analytic properties,
in \S 3, we will give the proof of the formula (1-7) or
Theorem 3-5.
As the notations becomes so complicated,
we will not mention the idea of our
derivation in detail here  but in Remark 3-6.
So we recommend that the reader should see the part in
Remark 3-7 (1) before
he starts to read \S 3.

Even though the notations becomes so complicated, the idea of our
approach is very {\it simple} as an analytic
process.

Before we finish the Introduction, we will comment on
another variant of a generalization of the elliptic
$\psi$ function and its recursion relation.
Recently Kanayama found the another recursion relation on
the function as a generalization of Eq.(1-1) to that of genus two,
$$
   \Phi_n :=\frac{\sigma( n u )}{\sigma( u)^{n^2}},
    \tag 1-12
$$
defined over the Jacobian \cite{\Kan}.
 $\Phi_n$ formally obeys the same as
Eq.(1-3) \cite{\Kan}. His approach is very resemble to ours.
 In order to compare
Kanayama's generalization \cite{\Kan} of Eq.(1-3) to genus
two case (defined over Jacobian $J$)  with Cantor's
one \cite{\Can} (defined over the curve $C$ itself),
I believe that our analytic approach is mathematically
important.

%\tvskip
%{\centerline{\bf{ Acknowledgment}}}
\tvskip

I'm deeply indebted to  Prof.~Y.~\^Onishi
for leading me this beautiful theory
of $\psi$-function, sending me
the work of Cantor and helpful discussions.
I thank Prof. V. Z. Enolskii for sending me
his interesting works  and helpful comments.
I am grateful to Prof. D. G. Cantor for critical
reading of this manuscript and several helpful
suggestions.

\tvskip
\centerline{\twobf \S 2 Abelian Functions of Genus Two}
\tvskip

This section reviews  theory of
hyperelliptic $\sigma$ and $\wp$ functions
\cite{\Bone, \Btwo, \Bthree}.
Recently  Buchstaber,  Enolskii, and Leykin
wrote a paper \cite{\BEL}
which contains very nice introduction
 to the theory  in terms of modern language;
it should be a guide to read this section.

In this article, we will deal with a hyperelliptic curve
$C$ of genus two, given by an affine equation,
$$
 \split
   y^2 &= f(x) \\
   &= \lambda_0 +\lambda_1 x
        +\lambda_2 x^2  +\lambda_3 x^3 +\lambda_4 x^4
+\lambda_5 x^{5},
\endsplit \tag 2-1
 $$
where $\lambda_{5}\equiv1$ and $\lambda_j$'s are
complex numbers.

\proclaim {\fp Definition 2-1 \cite{\Bone}, \cite{\Btwo},
\cite{\BEL\ Chapter 2}, \cite{\Oone\ p.384-386}}{\rm

 \roster

\item  Let us denote the homology of the hyperelliptic
curve $C$ by
$$
\roman{H}_1(X_g, \Bbb Z)
  =\Bbb Z\alpha_{1}\oplus\Bbb Z\beta_{1}
    \oplus\Bbb Z\alpha_{2}\oplus\Bbb Z\beta_{2},\tag 2-2
$$
where these intersections are given as
$[\alpha_i, \alpha_j]=0$, $[\beta_i, \beta_j]=0$ and
$[\alpha_i, \beta_j]=\delta_{i,j}$.

\item The unnormalized  differentials of first kind
are defined by,
$$   d u_1 := \frac{ d x}{2y}, \quad
      d u_2 :=  \frac{x d x}{2y}.\tag 2-3
$$

\item The unnormalized differentials of second kind are
defined by,
$$
     d r_{1}:=\dfrac{1}{2y}\left(
      \lambda_3 x+ 2\lambda_4 x^2+ 3\lambda_5 x^3 \right)dx ,
\quad
     d r_{2}:=\dfrac{1}{2y}
      \lambda_{5} x^2 dx .\tag 2-4
$$

\item The unnormalized period matrices are defined by,
$$   2 \pmb{\omega}':=\left[\matrix
        \dint_{\alpha_{1}}d u_{1} &
        \dint_{\alpha_{2}}d u_{1} \\
         \dint_{\alpha_{1}}d u_{2} &
         \dint_{\alpha_{2}}d u_{2}
         \endmatrix \right],
      2\pmb{\omega}'':=\left[\matrix
        \dint_{\beta_{1}}d u_{1} &
 \dint_{\beta_{2}}d u_{1} \\
         \dint_{\beta_{1}}d u_{2} &
 \dint_{\beta_{2}}d u_{2}
         \endmatrix \right],
 \quad
    \pmb{\omega}:=\left[\matrix \pmb{\omega}'
\\ \pmb{\omega}''
     \endmatrix\right].\tag 2-5
$$

\item The normalized period matrices are defined by,
$$
    \ ^t\left[\matrix d \hat u_{1}\\ d \hat u_{2}
        \endmatrix\right]
       :={\pmb{\omega}'}^{-1}  \ ^t\left[\matrix
          d u_{1} \\d u_{2}\endmatrix\right] ,\quad
   \pmb \tau:={\pmb{\omega}'}^{-1}\pmb{\omega}'',
   \quad
    \hat{\pmb{\omega}}:=\left[\matrix 1_g \\ \pmb \tau
     \endmatrix\right].\tag 2-6
$$

\item
We  define the Abel map for  symmetric product
of the curve $C$ {\it i.e.}, for points $\{ Q_1, Q_2 \}$
in the curve $C$ by,
$$       \hat w: \roman{Sym}^2(C) \longrightarrow
        \Bbb C^2, \quad
      \left( \hat w_k(Q_i):=\int_\infty^{Q_1}
      \hat d \hat u_k +\int_\infty^{Q_2}
      \hat d \hat u_k \right),
$$ $$ w:\roman{Sym}^2(C) \longrightarrow \Bbb C^2, \quad
      \left( w_k(Q_i):=\int_\infty^{Q_1}
      \hat d u_k +\int_\infty^{Q_2}
      \hat d u_k \right).
      \tag 2-7
$$
The Jacobian (Jacobi varieties) $\hat{\Cal J}$ and $\Cal J$
are defined as complex torus,
$$   \hat{\Cal J} := \Bbb C^2 /\hat{ \pmb{\Lambda}} ,
   \quad {\Cal J} := \Bbb C^2 /{ \pmb{\Lambda}} .
     \tag 2-8
$$
Here $\hat{ \pmb{\Lambda}}$ $({ \pmb{\Lambda}})$  is a
lattice generated by
$\hat{\pmb{\omega}}$ $({\pmb{\omega}})$.

\item The complete hyperelliptic integrals of the second kinds
are defined by
$$
   2 \pmb{\eta}':=\left[\matrix
        \dint_{\alpha_{1}}d r_{1} &
              \dint_{\alpha_{2}}d r_{1} \\
         \dint_{\alpha_{1}}d r_{2} &
                       \dint_{\alpha_{2}}d r_{2}
         \endmatrix \right],\quad
      2\pmb{\eta}'':=\left[\matrix
        \dint_{\beta_{1}}d r_{1} &
                      \dint_{\beta_{2}}d r_{1} \\
         \dint_{\beta_{1}}d r_{2} &
                      \dint_{\beta_{2}}d r_{2}
         \endmatrix \right].
      \tag 2-9
$$

\item We will define the Riemann theta function over
 $\Bbb C^2$
characterized by $\hat{ \pmb{\Lambda}}$,
$$
   \split
   \theta\negthinspace\left[\matrix a \\ b
      \endmatrix\right](z)
   & :=\theta\negthinspace\left[\matrix a \\ b
         \endmatrix\right]
     (z; \pmb\tau) \\
   & :=\sum_{n \in \Bbb Z^2} \exp \left[2\pi
     \sqrt{-1}\left\{
    \dfrac 12 \ ^t\negthinspace (n+a)\pmb \tau(n+a)
    + \ ^t\negthinspace (n+a)(z+b)\right\}\right],
 \endsplit
         \tag 2-10
$$
for $2$-dimensional vectors $a$ and $b$.

\endroster
}
\endproclaim

\tvskip
We will note that these contours in the integral are,
for example, given in \cite{\BEL\ p.19}. Thus above values can be
explicitly computed  for a given $y^2=f(x)$.

\vskip 0.5 cm
\proclaim {Definition 2-2 \cite{\Bone\ p.286},
\cite{B2\ p.336,\ p.353,\ p.370},
\cite{\BEL\ p.32,\ p.35}, \cite{\Oone\ p.386-7} }

{\rm
\roster
We will introduce the coordinate $(u_1, u_2)$ in $\Bbb C^2$
for  points $(x,y)$ and $(x_2,y_2)$
of the curve $y^2 = f(x)$
by $(u_1, u_2):=( w_1((x,y), (x_2,y_2)), w_2((x,y), (x_2,y_2)))$,
{\it i.e.},
$$
  u_j :=\int^{(x,y)}_\infty du_j
       +\int^{(x_2,y_2)}_\infty du_j .\tag 2-11
$$

\item The sigma function,
which is a homomorphic
function over $\Bbb C^2$, is defined by,
$$ \sigma(u):=\sigma(u;\pmb\omega):
  =\gamma \roman{exp}(-\dfrac{1}{2}\ ^t\ u
  \pmb\eta'{\pmb{\omega}'}^{-1}u)
  \vartheta\negthinspace
  \left[\matrix \delta'' \\ \delta' \endmatrix\right]
  ({(2\pmb{\omega}')}^{-1}u ;\pmb\tau) , \tag 2-12
$$
where $\gamma$ is a certain constant, which is determined
such that Proposition 2-4 holds, and,
$$
  \delta' :=\left[\matrix 1 \\ {}\\
       \dfrac {1}{2}\endmatrix\right],\quad
 \delta'':=\left[\matrix \dfrac{1}{2} \\
          \dfrac{1}{2} \endmatrix\right].
\tag 2-13
$$

  Let its derivative denote  $\sigma_\mu
:= \partial\sigma/\partial u_\mu$.

\item
The hyperelliptic $\wp$-function over
the Jacobian $\Cal J$ is defined by,
$$   \wp_{\mu\nu}(u):=-\dfrac{\partial^2}{\partial
   u_\mu\partial u_\nu}
   \log \sigma(u) . \tag 2-14
$$

\item
When $(x_2,y_2)$ is the infinity point, $(u_1,u_2)$ is
a function only of $(x,y)\in C$ and we refer
this $u$ by $u \in \iota(C)\subset \Cal J$;
the operation makes the curve $C$ embedded
into the Jacobian $\Cal J$, $\iota: C \hookrightarrow
\Cal J$.
\endroster
}
\endproclaim

\proclaim{\fp Proposition 2-3 }\it

Let us express
$\wp_{\mu\nu\rho}:=\partial \wp_{\mu\nu}(u)
  /\partial u_\rho$ and
$\wp_{\mu\nu\rho\lambda}:=\partial^2
 \wp_{\mu\nu}(u) /\partial u_\mu \partial u_\nu$.
Then hyperelliptic $\wp$-functions obey the relations,
$$
   \allowdisplaybreaks \align
     (H-1)\quad& \wp_{2222}-6\wp_{22}^2
      = 2 \lambda_3 \lambda_5
           + 4 \lambda_4 \wp_{22}
       + 4\lambda_5 \wp_{21} -12 \lambda_6 \wp_{11}
         ,\\
     (H-2)\quad& \wp_{2221}-6\wp_{22} \wp_{21}
      =
           4 \lambda_4 \wp_{21}
       - 2\lambda_5 \wp_{11}, \\
     (H-3)\quad& \wp_{2211}-4\wp_{21}^2
            -2\wp_{22} \wp_{11}
      =
          4 \lambda_2 \wp_{31}
       + 2\lambda_3 \wp_{21}, \\
     (H-4)\quad& \wp_{2111}-6\wp_{21} \wp_{11}
      =- 2\lambda_0 \lambda_5
      - 2 \lambda_1(\wp_{22})
       + 4\lambda_2 \wp_{21}, \\
     (H-5)\quad& \wp_{1111}-6\wp_{11}^2
      =- 4\lambda_0 \lambda_4+ 2 \lambda_1 \lambda_3
       -12\lambda_0( \wp_{22})
       + 4\lambda_1 \wp_{21} + 4 \lambda_2 \wp_{11}.
 \tag 2-15    \endalign
$$
$$   \allowdisplaybreaks \align
(I-0)\quad& \wp_{112}
      =\wp_{222}\wp_{12}+\wp_{122}\wp_{22}
         ,\\
     (I-1)\quad& \wp_{222}^2
      = 4(\wp_{22}^3 +\wp_{12}\wp_{22}+\lambda_4\wp_{22}^2+
            \wp_{11}+\lambda_3\wp_{22} +\lambda_2)
         ,\\
     (I-2)\quad& \wp_{222}\wp_{221}
      =4 (\wp_{12}\wp_{22}^2 -\frac{1}{2}(\wp_{11}\wp_{22}
             -\wp_{12}^2 + \lambda_3 \wp_{12}-\lambda_1) +
              \lambda_3\wp_{12} +\lambda_4 \wp_{12}\wp_{22})
               ,\\
     (I-3)\quad& \wp_{221}^2-
      =
      4 (\wp_{11}\wp_{22}^2 -
          (\wp_{11}\wp_{22} - \wp_{12}^2
           + \lambda_3 \wp_{12}-\lambda_1)\wp_{22}
         - \wp_{11}\wp_{12} + \lambda_4 \wp_{11}\wp_{22}\\
      &\quad   + \lambda_3 \wp_{12} \wp_{22}
   -\lambda_4 (\wp_{11}\wp_{22}-\wp_{12}^2 +\lambda_3\wp_{12}
       -\lambda_1) \\
    & \quad+ \lambda_4 \lambda_3 \wp_{12}-\lambda_1 \wp_{22}
      -\lambda_1\lambda_4+\lambda_0.
     \tag 2-16 \endalign
$$

\endproclaim

\demo{Proof}
See  p.155-6 in \cite{\Bthree}, p.86-88 in \cite{\BEL},
or \cite{\Gran}.\qed \enddemo

We also have the additional formula for them.

\proclaim{\fp Proposition 2-4 }\it

\roster

\item
$$
\frac{\sigma(u+v)\sigma(u-v)}{\sigma(u)^2\sigma(v)^2} = Q(u,v),
\tag 2-17
$$
where
$$
Q(u,v) := -(\wp_{11}(u)-\wp_{11}(v) + \wp_{12}(u)\wp_{22}(v) -
                          \wp_{12}(v)\wp_{22}(u)).
                          \tag 2-18
$$

\item
$$
\wp_{ij}(u+v) + \wp_{ij}(u-v ) = 2 \wp_{ij}(u)
  - \frac{Q_{ij} Q - Q_i Q_j}{Q^2}, \tag 2-19
$$
where
$ Q_{i}(u,v):=\partial_{u_i} Q$ and
$ Q_{ij}(u,v):=\partial_{u_j} Q_j$.

\endroster
\endproclaim

\demo{Proof}
See  p.372, p.381 in \cite{\B2} or  p.109-114 in
\cite{\BEL}.\qed\enddemo

\proclaim{\fp Proposition 2-5 }\it

In terms of $(x, y)$ and $(x_2,y_2)$ in Eq.(2-11), the
$\wp$ functions are expressed by
$$
        \wp_{12} = x x_2, \qquad \wp_{22} = x +x_2.
\tag 2-20
$$

\endproclaim

\demo{Proof} See  p.377 in \cite{\B2} or
 p.38, p.84 in \cite{\BEL}.\qed\enddemo

\vskip 0.5 cm
\centerline{\twobf \S 3. Recursion Relation
of Hyperelliptic $\psi$-function of Genus Two}
\vskip 0.5 cm

Let us define a hyperelliptic $\psi$ function of
genus two as
$$
        \psi_n(u)=\frac{\sigma( n u )}{\sigma_2( u)^{n^2}}, \tag 3-1
$$
for $u\in \iota (C)\subset\Cal J$.
In general, for $u\in \iota (C)\subset\Cal J$, $\sigma$ and
$\sigma_2(\equiv \partial\sigma/\partial u_2)$
vanish but by cancellation of zeros, the formula (3-1)
is well-defined and has a finite value.

We wish to derive the relation (1-7) over the embedded curve
$\iota(C)$ but the quantities and relations in the previous section
are defined over $\Cal J$.
Thus let us introduce a quantity,
$$
        \Psi_n(u)=\frac{\sigma( n u )}{\sigma_2( u)^{n^2}},
\tag 3-2
$$
for $u\in \Cal J$, which is limit as
$\psi_n(u) = \lim_{u\to \iota(C)} \Psi_n(u)$.
After considering  relations over $\Cal J$, we will restrict
some quantities into $\iota(C)$.
We also note that even if the local parameter $u$ is in
$\iota(C)$, the point $nu$ $(n>1)$ is in $\Cal J$
but not in $\iota(C)$ in general.

Using Eq.(3-2), the addition formula (2-17) is expressed by
 $$
 \frac{\Psi_{m+n}(u)\Psi_{m-n}(u)}
{\Psi_n(u)^2\Psi_m(u)^2}=Q(mu,nu). \tag 3-3
$$
For simplicity, we assume that $n$ and $m$ are
natural numbers grater than 1 and $u$ is a generic point in $\iota(C)$
hereafter,

\proclaim{\fp Lemma 3-1}\it

We introduce a quantity $\Xi_{3}(mu,u)$ as
$$
      \Xi_{3}(mu,u):= \frac{1}{\Psi_{m}(u)^3
           }\left(\Psi_{m-1}(u)^2 \Psi_{m+2}(u)
           +\Psi_{m+1}(u)^2 \Psi_{m-2}(u)\right),
       \tag 3-4
$$
for $u \in \Cal J$. Then $\Xi_3(mu,u)$ becomes,
$$
\split
      \Xi_{3}(mu,u)&=
\Psi_1(u)^6\Bigr( Q(mu,u)\bigr[2Q(mu,u)^2 +Q_{11}(mu,u)\\
&-\wp_{22}(u) Q_{12}(mu,u)+\wp_{12}(u)Q_{22}(mu,u)\bigr]\\
&\quad -  Q_{1}(mu,u)^2
-\wp_{22}(u)  Q_{1}(mu,u)Q_2(mu,u))\\
&+\wp_{12}(u) Q_{2}(mu,u)^2\Bigr).\\
\endsplit
\tag 3-5
$$
\endproclaim

\demo{Proof}
First we can rewrite $\Xi_3$ as,
$$
\split
      \Xi_{3}(mu,u)&= \Psi_1(u)^6
\frac{\Psi_{m-1}(u)^2 \Psi_{m+1}(u)^2}
 {\Psi_{m}(u)^4\Psi_{1}(u)^4}\left(\frac{\Psi_{m+2}(u) \Psi_{m}(u)}
 {\Psi_{m+1}(u)^2\Psi_{1}(u)^2}
           +\frac{\Psi_{m-2}(u) \Psi_{m}(u)}
 {\Psi_{m-1}(u)^2\Psi_{1}(u)^2}\right).
 \endsplit \tag 3-6
 $$
Using Proposition 2-4 (1) and (2) sequentially, it becomes
$$
 \split
&\Psi_1(u)^6 Q(mu,u)^2( Q((m+1)u,u) + Q((m-1)u, u) )\\
&=\Psi_1(u)^6 Q(mu,u)^2
\Bigr(2\wp_{11}(u) -2\wp_{11}(mu)\\
&+\frac{ Q_{11}(mu,u)Q(mu,u) -  Q_{1}(mu,u)^2}{Q(mu,u)^2}\\
&\quad -\wp_{22}(u)\Bigr(2\wp_{12}(mu)
-\frac{ Q_{12}(mu,u)Q(mu,u) -  Q_{1}(mu,u)Q_2(mu,u)}{Q(mu,u)^2}
\Bigr)\\
&\quad +\wp_{12}(u)\Bigr(2\wp_{22}(mu)
-\frac{ Q_{22}(mu,u)Q(mu,u) -  Q_{2}(mu,u)^2}{Q(mu,u)^2}
\Bigr)
\Bigr)\\
&=\Psi_1(u)^6\Bigr(2 Q(mu,u)^2
(\wp_{11}(u) -\wp_{11}(mu)+\wp_{22}(u)\wp_{12}(mu)
 -\wp_{12}(u)\wp_{22}(mu))\\
&\quad+Q_{11}(mu,u)Q(mu,u) -  Q_{1}(mu,u)^2\\
&\quad -\wp_{22}(u)( Q_{12}(mu,u)Q(mu,u) -  Q_{1}(mu,u)Q_2(mu,u))\\
&\quad+\wp_{12}(u)( Q_{22}(mu,u)Q(mu,u) -  Q_{2}(mu,u)^2)
\Bigr).
\endsplit\tag 3-7
$$
By arranging these terms, we obtain Eq.(3-5).
\qed\enddemo

\vskip 0.5 cm

Hereafter,
let us consider the restriction of these quantities to
the embedded curve $\iota(C) \subset \Cal J$.

\proclaim{\fp Lemma 3-2}\it

The following relations hold over $\iota(C)$:

\roster
\item For $u\in \iota(C)$,
$$
        \wp_{12}(2 u) = -x^2, \quad \wp_{22}(2 u) = 2x.
        \tag 3-8
$$

\item For $u\in \iota(C)$,
$$
         \quad \psi_2(u) = 2y. \tag 3-9
$$

\item
$$
        \lim_{u \to \iota (C)} \left(\Psi_1(u)^2
               \wp_{11 }(u)\right)=x^2,
$$
$$
        \lim_{u \to \iota (C)} \left(\Psi_1(u)^2
               \wp_{12 }(u)\right)=-x,
$$
$$
        \lim_{u \to \iota (C)} \left(\Psi_1(u)^2
                 \wp_{22 }(u)\right)=1.
         \tag 3-10
$$

\endroster
\endproclaim

\demo{Proof}
Proposition 2-5 leads (1) directly. On (2),  $\psi_2(u)
=\dfrac{ \sigma(2u)}{\sigma_2(u)^4}$ was given in
\cite{\Oone}. From Proposition 2-5, we have
$$
        x =\lim_{u \to \iota(C)} \frac{\wp_{12}(u)}{\wp_{22}(u)}=
        \lim_{u \to \iota(C)} \frac{\sigma_{1}(u)}{\sigma_{2}(u)}.
    \tag 3-11
$$
Noting the definition of $\wp$ and using the relation,
$$
        \lim_{u \to \iota (C)} \left(\Psi_1(u)^2
          \wp_{i j }(u)\right)
                       = \lim_{u \to \iota (C)}
 \left( \frac{\sigma(u)^2}{\sigma_2(u)^2}
        \frac{\sigma_{i}(u)\sigma_j(u) -
\sigma_{ij }(u)\sigma(u)}
       {\sigma(u)^2}\right), \tag 3-12
$$
(3) is obtained.\qed
\enddemo

\proclaim{\fp Lemma 3-3}\it

\roster

\item
$$
        \lim_{u \to \iota (C)}Q(mu,2u ) =
          (\wp_{11}(mu) -\wp_{11}(2u) -2 \wp_{12}(mu)x
             - \wp_{22}(mu)x^2). \tag 3-13
$$

\item
The quantities
$$
q(mu,u):= \lim_{u \to \iota (C)} \Psi_1(u)^2 Q(mu,u),\quad
q_i(mu,u):= \lim_{u \to \iota (C)} \Psi_1(u)^2 Q_i(m u,u),
$$
$$
q_{ij}(mu,u):= \lim_{u \to \iota (C)} \Psi_1(u)^2 Q_{ij}(mu,u),
\tag 3-14
$$
become
$$
   \split
        q(mu,u) &= -x^2 + \wp_{12}(mu)+\wp_{22} x,\\
        q_{ij}(mu,u)&=\wp_{12ij}(mu) + \wp_{22ij}(mu)x,\\
q_{i}(mu,u)&=\wp_{12i}(mu) + \wp_{22i}(mu)x.
\endsplit\tag 3-15
$$

\item
By letting $\xi_3(mu,u):=\lim_{u\in \iota(C)} \Xi_3(mu,u)$,
$$
\split
        \xi_3(mu,u)&= q(mu,u)\Bigr(2 q(mu,u)^2 -
 q_{12}(mu,u)-xq_{22}(mu,u)\Bigr)\\
&\quad -
 q_{2}(mu,u)\Bigr(q_1(mu,u))+xq_2(mu,u) )\Bigr).
\endsplit\tag 3-16
$$

\endroster
\endproclaim

\demo{Proof} Addition formula (2-17) of $Q(mu,2u)$
 is restricted to $\iota(C)$
noting Lemma 3-2 (3), and then (1) is obtained. Similarly
(2) and (3) are proved. In (3), we note that
$$
   \lim_{u \to \iota (C)} \Psi_1(u)^6 Q_{11}(mu,u)^2 =0,
        \quad
        \lim_{u \to \iota (C)} \Psi_1(u)^6 Q(mu,u)Q_{11}(mu,u) =0,
        \tag 3-17
$$
due to the order of the zero at $\iota (C)$.
\qed \enddemo

\proclaim{\fp Lemma 3-4}\it

For $u\in \iota (C)$,
we introduce the quantities,

$$
        \xi_0(mu, nu):=
             \psi_2(u)^2 \dfrac{\psi_{m-n}(u)\psi_{m+n}(u)}
                {\psi_m(u)^2\psi_n(u)^2},
\quad
        \xi_1(mu):= \frac{\psi_{m-2}(u)
              \psi_{m+2}(u)}{\psi_m(u)^2},
$$
$$
\split
        \xi_2(mu,nu)&:=\frac{\psi_{m-1}(u)\psi_{m+1}(u)}
{\psi_{m}(u)^2\psi_{n}(u)^3}(\psi_{n-1}(u)^2 \psi_{n+2}(u)
           +\psi_{n+1}(u)^2 \psi_{n-2}(u)) \\
&-\frac{\psi_{n-1}(u)\psi_{n+1}(u)}
{\psi_{n}(u)^2\psi_{m}(u)^3}(\psi_{m-1}(u)^2 \psi_{m+2}(u)
           +\psi_{m+1}(u)^2 \psi_{m-2}(u)).
\endsplit \tag 3-18
$$
They satisfy the following relations:

\roster

\item
$$
        \xi_0(mu, nu)=-4y^2
 (\wp_{11}(mu) -\wp_{11}(nu)
  -2 \wp_{12}(mu)\wp_{12}(nu) - \wp_{22}(mu) \wp_{22}(nu)).
    \tag 3-19
$$

\item
$$
        \xi_1(mu)=4y^2
 (\wp_{11}(mu) -\wp_{11}(2u) -2 \wp_{12}(mu)x
           - \wp_{22}(mu)x^2).
\tag 3-20
$$

\item
$$
\split
        \xi_2(mu,nu)
         &= 4 y^2( \psi_{22}(mu)x^2+2\wp_{12}(mu)x
           -\wp_{12}(nu)\wp_{22}(mu)\\
        & \quad
      - \psi_{22}(nu)x^2-2\wp_{12}(nu)x
           -\wp_{12}(mu)\wp_{22}(nu)).
\endsplit\tag 3-21
$$

\endroster
\endproclaim

\demo{Proof}
(1) and (2) are obvious by Eq.(3-3) and Proposition 2-4.
 (3) needs precise computations.
$\xi_2(mu,nu)$ consists of $\xi_3(mu,u)$ and $\xi_3(nu,u)$
due to the definition in Lemma 3-3,
$$
\split
        \xi_2(mu,nu)&=\frac{\psi_{m-1}(u)\psi_{m+1}(u)}
{\psi_{m}(u)^2}\xi_3(nu,u)
-\frac{\psi_{n-1}(u)\psi_{n+1}(u)}
{\psi_{r}(u)^2}\xi_3(mu,u)\\
&=
( x^2 - \psi_{22}(nu)x -\psi_{12}(nu))\xi_{3}(mu,u)\\
&\quad\quad         -   ( x^2 - \psi_{22}(mu)x
            -\psi_{12}(mu))\xi_{3}(nu,u).
\endsplit\tag 3-22
$$
Here we use Eq.(2-18), Eq.(3-3) and Eq.(3-15).
Let us consider the parts of $\xi_{3}(mu,u)$
in Eq.(3-16),
$$
\split
        q_{12}(mu,u)
&= 6\wp_{12}(mu)\wp_{22}(mu) - 4(\wp_{11}(mu)\wp_{22}(mu)
     -\wp_{12}(mu)^2)
    + 2 \lambda_2 \wp_{12}(mu)\\
       &  + x ( 6 \wp_{12}(mu)\wp_{22}(mu) - 2\wp_{11}(mu)
 + 4 \lambda_1 \wp_{12}(mu)),\\
        q_{22}(mu,u) &= 6\wp_{12}(mu)\wp_{22}(mu)
 - 2\wp_{11}(mu)+4\lambda_1\wp_{12}(mu)\\
       &  + x ( 6 \wp_{22}(mu)^2 +4\wp_{12}(mu)
+ 4 \lambda_1 \wp_{22}(mu)+ 2\lambda_2),
\endsplit
$$
$$
        q_1(mu,u) = \wp_{112}(mu) + x\wp_{221}(mu),
       \quad q_2(mu,u) = \wp_{122}(mu) + x\wp_{222}(mu).
\tag 3-23
$$
By using above parts,
the constituent of the first term in $\xi_3(mu,u)$ is expressed by
$$
\split
2q(mu,u)^2 &-
 q_{12}(mu,u)-xq_{22}(mu,u)\\
&=2 x^4 - 4 x^3\wp_{22}(mu)\\
& - (4\wp_{22}(mu)^2 + 8 \wp_{12}(mu)+4\lambda_1 \wp_{22}(mu)
   +2\lambda_2) x^2\\
&- (8\wp_{12}(mu)\wp_{22}(mu) - 4\wp_{11}(mu)+8\lambda_1
       \wp_{12}(mu) ) x\\
&- (2\wp_{11}(mu)\wp_{22}(mu) + 2\wp_{12}(mu)^2
    +2\lambda_2 \wp_{12}(mu) ),
\endsplit\tag 3-24
$$
and the second term in $\xi_3(mu,u)$ becomes
$$
\split
q_{2}(mu,u)&\Bigr(q_1(mu,u))+xq_2(mu,u) )\Bigr)\\
&=( \wp_{122}(mu) + x\wp_{222}(mu))
      ( \wp_{122}(mu) +2 x\wp_{221}(mu)+ x^2\wp_{222}(mu))\\
&=\wp_{222}(mu)^2 x^3 + 3 \wp_{122}(mu)\wp_{222}(mu) x^2\\
&+(\wp_{12}(mu)\wp_{222}(mu)^2
- \wp_{22}(mu) \wp_{122}(mu)\wp_{222}(mu)+2\wp_{122}(mu)^2)x\\
 & -\wp_{22}(mu)\wp_{122}(mu)^2
- \wp_{12}(mu) \wp_{122}(mu)\wp_{222}(mu).
\endsplit\tag 3-25
$$
Using the relation (I-0)-(I-3) in Proposition 2-3,
$q_{2}(mu,u)\Bigr(q_1(mu,u))+xq_2(mu,u) )\Bigr)$ is reduced
to the formula consisting only of $\wp_{ij}(mu)$. After tedious
computations, we have the relation $\xi_2(mu,nu)$.
\qed\enddemo

Since
$
        \xi_0(nu,mu)+ \xi_1(mu)-\xi_1(nu) - \xi_2(mu,nu) =0
$,
we have a part of
the recursion relation of $\psi$-function of genus
two. From  the definition of $\psi_n$, we have the relations,
$\psi_0\equiv \psi_1 \equiv0$ for $u\in \iota(C)$
\cite{\Oone, \Otwo},
the recursion relation can be
extended to all  integers $n$ and $m$ ($m\ge n\ge 0$).

\proclaim{\fp Theorem 3-5}\it

For integers $m$, $n$, ($m \ge n \ge 0$),
$$
  \psi_2\psi_2\psi_m\psi_n \psi_{n+m} \psi_{m-n}
=\left| \matrix
 \psi_{m-2}\psi_n & \psi_{m-1}\psi_{n+1} & \psi_m\psi_{n+2}\\
\psi_{m-1}\psi_{n-1} & \psi_m\psi_n & \psi_{m+1}\psi_{n+1}\\
\psi_m\psi_{n-2} &\psi_{m+1}\psi_{n-1}& \psi_{m+2}\psi_n
\endmatrix \right|.  \tag 3-26
$$

\endproclaim

\tvskip
\proclaim{\fp Remark 3-6}

{\rm
\roster

\item
As mentioned in Introduction,
 let us summarize this article here.
In fact
we derived the recursion relation (3-26)
 but the process we stepped
is too confused and looks haphazard.
Thus we will mention the idea
of our derivation and recapitulate the process here.

In our proofs of the relations in Theorem 3-5 and
Eq.(1-3), which are cases of genera one and two, addition
formulae play important roles. Especially, linear terms of
$\wp$-function in the addition formulae are essential.
For the case of genus one, the role of the linear term
with respect to $\wp$ was mentioned in Eq.(1-6).
Even for the case of genus two, the idea to derive
the recursion relation (3-26) is the same.

As we have the addition formula of $\wp$
in Eq.(1-10) or Eqs.(2-17) and (2-18),
the factor $\sigma(u+v)\sigma(u-v)$ in the left hand
side of  Eq.(2-17) is an essential part of $\psi_{m+n}\psi_{m-n}$
in the left hand side of Eq.(3-26) since $\psi$ is
defined as Eq.(3-1) in terms of the $\sigma$ function.
Combination of the linear terms with respect to $\wp$'s
in Eq.(2-18) can be canceled
like the genus one case. Thus by introducing  $\xi_0$
and $\xi_1$ in Eq.(3-18),
we considered $\xi_0(nu,mu)+ \xi_1(mu)-\xi_1(nu)$ like Eq.(1-6);
from Eqs.(3-19) and (3-20) in Lemma 3-4, these linear terms
without $x$ nor $x^2$ vanishes.
Identification $2y$ with $\psi_2$ in
Eq.(3-9) roughly gives a recursion relation.
However in the addition formula Eq.(2-18), there appear
the quadratic terms on $\wp$'s and thus we need some technical
manipulations. $\xi_2$ in Eq.(3-18) appears to
remove the excess terms in
$\xi_0(nu,mu)+ \xi_1(mu)-\xi_1(nu)$ due to the formula (3-21).

Further $\psi$ function of genus two is defined over
a curve itself $\iota(C)$
rather than the Jacobian $\Cal J$ but
the $\wp$'s and $\sigma$'s are defined over the
Jacobian $\Cal J$.  Hence in order to perform the
process mentioned above, we need consider the restriction
of the domain of the functions and the relations.
We started with functions over the Jacobian $\Cal J$,
{\it i.e.} $\Psi$ in Eq.(3-2), $\Xi$'s in Eq.(3-4), $Q$ in
Eq.(2-17) and so on,
and then investigated them. After restricting their domain
to $\iota(C)$, we obtained Theorem 3-5.

Accordingly we emphasize again that the idea of our derivations
of the recursion relations of genera one and two is
very simple and in the derivation, linear terms of
$\wp$-function in the addition formulae play essential roles.

\item
After seeing the success of the idea of the proofs of genera
one and two,  one might  consider more general relation
for higher genus. However he must encounter the difficulty.
Because in the addition formula for the case of genus three,
$$
\split
-\frac{\sigma(u+v)\sigma(u-v)}{\sigma(u)^2\sigma(v)^2}& =
(\wp_{31}(u)-\wp_{31}(v))^2 -
( \wp_{31}(u)-\wp_{31}(v))( \wp_{22}(u)-\wp_{22}(v)) \\
&-( \wp_{33}(u)-\wp_{33}(v))( \wp_{11}(u)-\wp_{11}(v)) \\
&+( \wp_{12}(u)-\wp_{12}(v))( \wp_{23}(u)-\wp_{23}(v)). \\
\endsplit \tag 3-27
$$
there is no linear term with respect to $\wp$
for the hyperelliptic $\sigma$ and $\wp$ functions of
genus three.
The $\psi$ function of genus three is also given
by
$$
        \psi_n(u)=\frac{\sigma( n u )}{\sigma_2( u)^{n^2}},
\tag 3-28
$$
for a coordinate $u$ of the embedded hyperelliptic
curve in its related Jacobian of genus three \cite{\Othree}.
In other words, the attempt of the
extension of the derivation Eq.(1-3)-(1-6) might fail
for the case of genus three.
In order to reveal the difference between the properties
of $\psi$ functions of the cases $g\le 2$ and $g= 3$,
we  believe that our approach is also mathematically
important.

}
\endproclaim

\enddocument

\enddocument